\documentclass{iopart}  

\usepackage{cite}
\usepackage[]{graphicx}
\usepackage[percent]{overpic}
\usepackage[]{amssymb}
\usepackage{hyperref}
\begin{document}

\title[Necessary and sufficient conditions for reflectionless transformation media]{Necessary and sufficient conditions for reflectionless transformation media in an isotropic and homogenous background}

\author{Wei Yan, Min Yan, Min Qiu$^*$}%

\address{Laboratory of Optics, Photonics and Quantum Electronics,
  Department of Microelectronics and Applied Physics, Royal Institute
  of Technology, 164 40 Kista, Sweden}%
\ead{min@kth.se}


\begin{abstract}
It has been known that, keeping the outer boundary coordinates
intact before and after a coordinate transformation is a sufficient
condition for obtaining a reflectionless transformation medium. Here
we prove that it is also a necessary condition for reflectionless
transformation media in an isotropic and homogenous background. Our
analytical results show that the outer boundary coordinates of a
reflectionless transformation medium must be the same as the
original coordinates with a combination of rotation and
displacement, which is equivalent to situation that the boundary
coordinates are kept intact before and after transformation.
\end{abstract}

\maketitle

\section{Introduction}
Recently, invisibility cloaks have received a lot of attentions
\cite{Pendry}-\cite{U2}, due to their amazing properties of
excluding light from a protected object without perturbing exterior
fields. The real breakthrough behind the story is the theoretical
tools used to construct the cloak: transformation optics, in which
coordinate transformation is applied to obtain a new set of
permittivity and permeability tensors whereas Maxwell's equations
are still form invariant. Inspired by invisibility cloaks, different
kinds of media constructed by the coordinate transformation method
are proposed to control electromagnetic fields for novel
applications, such as field concentrator \cite{Rahm}, field shifter
and splitter \cite{Rahm2}, field rotator \cite{Ychen},
electromagnetic wormholes \cite{Greenleaf}, and cylindrical
superlens \cite{MY}. Such media obtained from the coordinate
transformation are called as transformation media, in which the
behaviors of light are determined by coordinate transformation
functions.

For most of applications, it is desirable to have transformation
media that are reflectionless to external light. It has been known
that \cite{Pendry,Ychen,Yan}, keeping the outer boundary coordinates
intact before and after a coordinate transformation is a sufficient
condition for obtaining a reflectionless transformation medium. A
strict proof has been given in Ref. \cite{Yan}. However, it is still
unclear that whether the condition is the only necessary condition
for reflectionless transformation media, and whether other
sufficient conditions exist for reflectionless transformation media.
In the present paper, we analytically present the necessary and
sufficient condition for reflectionless transformation media, in
particular, in an isotropic and homogenous background medium. Some
numerical examples are also given to confirm our findings.

\section{Transformation media}
Maxwell's Equations without sources take the form as $\nabla  \times
{\bf E} =  - \frac{{\partial {\bf B}}}{{\partial t}}$, $\nabla
\times {\bf H} = \frac{{\partial {\bf D}}}{{\partial t}}$, $\nabla
\cdot {\bf D} = 0$, $\nabla \cdot {\bf B} = 0$, with ${\bf D} =
\overline{\overline \varepsilon }  \cdot {\bf E}$, ${\bf B} =
\overline{\overline \mu}  \cdot{\bf H}$. Consider the transformation
from the Cartesian coordinate system $(x,y,z)$ to an arbitrary
curved coordinate system $( q_1,q_2,q_3)$ (illustrated in Fig. 1),
which can be described by a set of equations $x = f_1 (q_1 ,q_{2,}
q_3 )$, $y = f_2 (q_1 ,q_{2,} q_3 )$, $z = f_3 (q_1 ,q_{2,} q_3)$.
Then Maxwell's equations in the curved coordinate system can be
expressed as $\nabla _q  \times \widehat {\bf E} =-\frac{{\partial
\widehat {\bf B}}}{{\partial t}}$, $\nabla _q \times \widehat {\bf
H} = \frac{{\partial \widehat {\bf D}}}{{\partial t}}$, $\nabla_q
\cdot \widehat {\bf D} = 0$, $\nabla_q \cdot \widehat {\bf B} = 0$,
with $\widehat {\bf D} =\widehat {\overline{\overline \varepsilon }
} \cdot \widehat {\bf E}$, $\widehat {\bf B} = \widehat
{\overline{\overline \mu} } \cdot \widehat {\bf H}$, where
\begin{equation}
\widehat{\overline{\overline \varepsilon } } = {{ \det (g)}({g^T})^{
- 1} } {\overline{\overline \varepsilon } }g^{-1} ,\quad
\widehat{\overline{\overline \mu } } = {{ \det (g)}({g^T})^{ - 1}
}\widehat {\overline{\overline \mu } }g^{-1},
\end{equation}
\begin{equation}
\widehat {\bf E} = g{\bf E},\quad \widehat {\bf H} = g{\bf H},
\end{equation}
with
\begin{eqnarray}
g = \left[ {\begin{array}{*{20}c}
   {\frac{{\partial f_1 }}{{\partial q_1 }}} & {\frac{{\partial f_2 }}{{\partial q_1 }}} & {\frac{{\partial f_3 }}{{\partial q_1 }}}  \\
   {\frac{{\partial f_1 }}{{\partial q_2 }}} & {\frac{{\partial f_2 }}{{\partial q_2 }}} & {\frac{{\partial f_3 }}{{\partial q_2 }}}  \\
   {\frac{{\partial f_1 }}{{\partial q_3 }}} & {\frac{{\partial f_2 }}{{\partial q_3 }}} & {\frac{{\partial f_3 }}{{\partial q_3 }}}  \\
\end{array}} \right].
\end{eqnarray}

\begin{figure}[htbp]
\centering
\includegraphics[width=14cm]{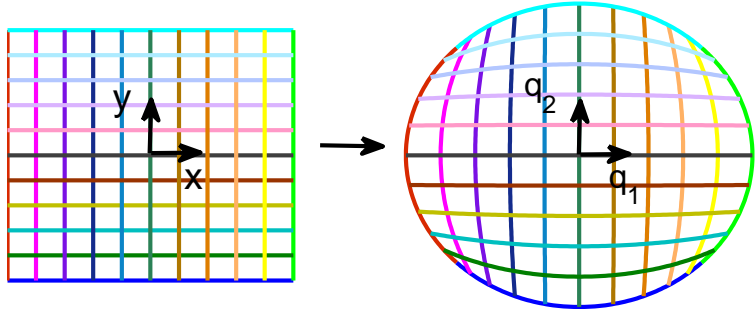}
\caption{Illustration of coordinate transformation.}
\end{figure}

It is seen that Maxwell's equations have the same form in any curved
coordinate system. However, the permittivity and permeability change
their values, which are expressed as the functions of
$(q_1,q_2,q_3)$ in Eq. 1. The media with such permittivity and
permeability are thus transformation media, where $(q_1,q_2,q_3)$
are treated as cartesian coordinates in physical space.

\section{Necessary condition for reflectionless transformation media}
Consider a transformation medium obtained by a coordinate
transformation from a homogenous and isotropic background (e.g.
vacuum) with permittivity $\epsilon$ and permeability $\mu$, and
placed in the same medium.

To derive the necessary condition for reflectionless transformation
media, we take the reflectionless property as a prerequisite. Since
the incident waves can always be expressed as the superposition of
the eigen waves, no reflection should occur for each eigen wave
incidence. It is known that the eigen wave in a homogenous and
isotropic medium is a plane wave. Consider an arbitrary propagating
plane wave incident upon the transformation media. The boundary
interacting with the incident wave between the transformation medium
and surrounding medium (i.e., the original medium) is denoted by
$S$. The electric and magnetic fields of the incident plane wave are
denoted by
\begin{equation}
{\bf E}_i(x,y,z)  = \overline {\bf E}_i\exp (i{\bf k}_i \cdot {\bf
r}),\,\, {\bf H}_i(x,y,z) = \overline {\bf H}_i\exp (i{\bf k}_i
\cdot {\bf r}),
\end{equation}
where $k_i  = \omega \sqrt {\mu \varepsilon}$; $r=q_1\widehat {\bf
x}+q_2\widehat {\bf y}+q_3\widehat {\bf z}$; $\overline {\bf E}_i$
and $\overline {\bf H}_i$ are constant vectors, representing the
directions and magnitudes of electric and magnetic fields,
respectively; $|\overline {\bf E}_i|/|\overline {\bf H}_i|=\sqrt{\mu
/ \varepsilon }$. Since the eigen waves in the transformation medium
relate to those in the original medium, i.e., surrounding medium, by
Eq. 2, the transmitted fields in the transformation medium can
always be expressed as
\begin{equation}
\fl{\bf E}_t(q_1,q_2,q_3) = \sum\limits_{n = - \infty }^{ + \infty }
{g \overline {\bf E}_n\exp (i{\bf k}_n \cdot {\bf r}^{'} ) },\,\,
{\bf H}_t(q_1,q_2,q) = \sum\limits_{n = - \infty }^{ + \infty }
{g\overline {\bf H}_n \exp (i{\bf k}_n \cdot {\bf r}^{'} )},
\end{equation}
where $ r^{'} = f_1 (q_{1,} q_{2,} q_3 )\widehat {\bf x} + f_2
(q_{1,} q_{2,} q_3 )\widehat {\bf y} + f_3 (q_{1,} q_{2,} q_3
)\widehat {\bf z}$; $\overline {\bf E}_n\exp (i{\bf k}_n \cdot {\bf
r} )$ and $\overline {\bf H}_n\exp (i{\bf k}_n \cdot {\bf r} )$
represent the eigen plane waves in the surrounding medium, including
both propagating and evanescent components.

Eq. 5 can also be rewritten as ${\bf E}_t(q_1,q_2,q_3) = \exp (i{\bf
k}_i \cdot {\bf r} )\sum\limits_{n = - \infty }^{ + \infty } {g
\overline {\bf E}_n\exp [i({\bf k}_n \cdot {\bf r}^{'}-{\bf k}_i
\cdot {\bf r}) ] }$, ${\bf H}_t(q_1,q_2,q_3) = \exp (i{\bf k}_i
\cdot {\bf r} )\sum\limits_{n = - \infty }^{ + \infty } {g\overline
{\bf H}_n \exp [i({\bf k}_n \cdot {\bf r}^{'}- {\bf k}_i \cdot {\bf
r})]}$. Boundary conditions for no reflection requires that the
tangential components of ${\bf E}_i$ (${\bf H}_i$) and ${\bf E}_t$
(${\bf H}_t$) keep continuous across the boundary. To satisfy the
boundary conditions, the phase factor of ${\bf E}_i$ and ${\bf H}_i$
at the boundary $S$, i.e., $\exp (i{\bf k}_i \cdot {\bf r_S})$,
should be matched with the the phase factor of ${\bf E}_t$ and ${\bf
H}_t$ at the same boundary. (Here the subscript $S$ indicates the
values at the boundary.) Thus, ${\bf k}_n \cdot {\bf r_S}^{'}-{\bf
k}_i \cdot {\bf r_S}$ should be a constant at any points of the
boundary $S$. While since ${\bf k}_n$ could have an arbitrary
direction, at most one of ${\bf k}_n \cdot {\bf r_s}^{'}-{\bf k}_i
\cdot {\bf r_s}$ can be constant. It can then be concluded that only
one term of $\overline {\bf E}_n$ ($\overline {\bf H}_n$) is
non-zero, denoted by $\overline {\bf E}_t$ ($\overline {\bf H}_t$),
while the other terms are all zeros. Now we have ${\bf E}_t =
g\overline {\bf E}_t \exp (i{\bf k}_t \cdot {\bf r}^{'}),\,\, {\bf
H}_t(x,y,z) = g\overline {\bf H}_t\exp (i{\bf k}_t \cdot {\bf
r}^{'})$. It indicates that only one eigen wave is excited in the
transformation medium for one specific incident eigen wave. The
boundary conditions can then be written as
\begin{equation}
\widehat {\bf t} \cdot \overline {\bf E}_i \exp (i{\bf k}_i \cdot
{\bf r_S}) = \widehat {\bf t} \cdot g_S\overline{\bf E}_t \exp
(i{\bf k}_t \cdot {\bf r_S}^{'} ),
\end{equation}
\begin{equation}
\widehat {\bf t} \cdot \overline{\bf H}_i \exp (i{\bf k}_i \cdot
{\bf r_S}) = \widehat {\bf t} \cdot g_S\overline{\bf H}_t \exp
(i{\bf k}_t \cdot {\bf r_S}^{'} ),
\end{equation}
where $\widehat {\bf t}$ represents the unit vector in the
tangential direction of the boundary $S$. The phase matching
requires that
\begin{equation}
{\bf k}_i  \cdot {\bf r_S} = {\bf k}_t  \cdot {\bf r_S}^{'}  + \phi,
\end{equation}
where $\phi$ is a constant.

Since the incident eigen wave $\{\overline{\bf E}_i \exp (i{\bf
k}_i\cdot {\bf r}), \overline{\bf H}_i \exp (i{\bf k}_i\cdot {\bf
r})\}$ is a propagating plane wave, $\{\overline{\bf E}_t \exp
(i{\bf k}_t\cdot {\bf r}), \overline{\bf H}_t \exp (i{\bf k}_t\cdot
{\bf r})\}$ should also be the propagating plane wave for the
conservation of energy through the boundary. The unit vectors in the
directions of $\overline{\bf E}_t$, $\overline{\bf H}_t$, and ${\bf
k}_t$ are denoted by $\widehat {\bf e}_t^1$, $\widehat {\bf e}_t^2$
and $\widehat {\bf e}_t^3$, which are orthogonal to each other.
While the unit vectors in the directions of $\overline{\bf E}_i$,
$\overline{\bf H}_i$, and ${\bf k}_i$ are denoted by $\widehat {\bf
e}_i^1$, $\widehat {\bf e}_i^2$ and $\widehat {\bf e}_i^3$, which
are also orthogonal to each other. It is obvious that $({\widehat
{\bf e}_t^1, \widehat {\bf e}_t^2,\widehat {\bf e}_t^3})$ can be
coincident with $({\widehat {\bf e}_i^1, \widehat {\bf
e}_i^2,\widehat {\bf e}_i^3})$ after a certain rotation, where a
${3\times3}$ orthogonal matrix $R$ can be defined to characterize
this rotation, with ${\bf e}_t^j=R{\bf e}_i^j$ $(j=1,2,3)$.
Therefore, ${\bf E}_t$ and ${\bf H}_t$ can be expressed as
\begin{equation}
\fl \;\;\;\;\;\;\;\;\;\;\;{\bf E}_t = gR^{-1}q\overline {\bf E}_i
\exp (i{\bf k}_i \cdot R{\bf r}^{'}),\,\, {\bf H}_t(x,y,z)
=gR^{-1}q\overline {\bf H}_i\exp (i{\bf k}_i \cdot R{\bf r}^{'}),
\end{equation}
where $q$ is a constant, and $R^{-1}$ is the inversion of $R$. The
phase matching condition at the boundary expressed in Eq. 8 can thus
be written as
\begin{equation}
{\bf k}_i  \cdot {\bf r_S} ={\bf k}_i  \cdot R{\bf r_S}^{'}+ \phi,
\end{equation}
which requires ${\bf r_s}-R{\bf r_s}^{'}$ to be a constant vector,
i.e.,
\begin{equation}
{\bf r_S}=R{\bf r_S}^{'}+{\bf u}, \label{reflectionless}
\end{equation}
where ${\bf u}$ denoted by ${\bf u}=u_1 \widehat {\bf x}+u_2
\widehat {\bf y}+u_3 \widehat {\bf z}$ is a constant vector, and
$\phi={\bf k}_i\cdot {\bf u}$. Eq.~\ref{reflectionless} is the
necessary condition for reflectionless transformation media,
resulted from the requirement of phase matching at the boundary.

\section{Sufficient condition for reflectionless transformation media}
In this section, we briefly prove that Eq. 11 is also the sufficient
condition for reflectionless transformation media. Consider a
propagating plane wave ${\bf E}_i$ and ${\bf H}_i$ expressed in Eq.
(4) incident upon the transformation medium, whose outer boundary
coordinates $\bf r_S$ are related to the transformed coordinates
$\bf r^{'}_S$ by Eq. 11. The the transmitted fields ${\bf E}_t$ and
${\bf H}_t$ in the transformation medium are directly obtained by
the coordinate transformation through Eq. 9, which satisfy Maxwell's
equations in the transformation medium. Below we will show that the
tangential fields of the incident wave in Eq. 4 equal to the
tangential fields of the transmitted fields in Eq. 9, which means
the transformation medium is reflectionless.

From Eq. 11, we can obtain ${\bf r_S}^{'}=R^{-1}({\bf r_S}-{\bf
u})=R^{T}({\bf r_S}-{\bf u})$, where the superscript "$T$" denotes
the transpose of matrix. Here, it is noted that $R^{-1}=R^{T}$ since
$R$ is an orthogonal matrix. Because $ {\bf r_s}^{'} = f_1 (q_{1,}
q_{2,} q_3 )\widehat {\bf x} + f_2 (q_{1,} q_{2,} q_3 )\widehat {\bf
y} + f_3 (q_{1,} q_{2,} q_3 )\widehat {\bf z}$, we have
$f_j(q_1,q_2,q_3)=R_{1j}(q_1-u_1)+R_{2j}(q_2-u_2)+R_{3j}(q_3-u_3)$
$(j=1,2,3)$. It is obvious that $\nabla_q f_j- \nabla_q
[R_{1j}(q_1-u_1)+R_{2j}(q_2-u_2)+R_{3j}(q_3-u_3)]$, i.e., $(\partial
f_j /\partial q_1  - R_{1j} )\widehat {\bf x} + (\partial f_j
/\partial q_2  - R_{2j} )\widehat {\bf y} + (\partial f_j /\partial
q_3 - R_{3j} )\widehat {\bf z}$, lie in the same line as the normal
direction of $S$. Therefore, $g_S$ can be expressed as
\begin{equation}
g_S=[F_1{\widehat {\bf n}},\; F_2\widehat {\bf n},\; F_3\widehat
{\bf n}]+R,
\end{equation}
where $\widehat {\bf n}$ is the unit vector in the normal direction
of the boundary $S$; $F_j \widehat {\bf n}=(\partial f_j /\partial
q_1 - R_{1j} )\widehat {\bf x} + (\partial f_j /\partial q_2  -
R_{2j} )\widehat {\bf y} + (\partial f_j /\partial q_3  - R_{3j}
)\widehat {\bf z}$. Then, we obtain that $g_SR^{-1}=[F_1{\widehat
{\bf n}},\; F_2\widehat {\bf n},\; F_3\widehat {\bf
n}]R^{-1}+diag[1,1,1]$, and
\begin{equation}
\widehat {\bf t}\cdot g_sR^{-1}=\widehat {\bf t}.
\end{equation}
Observing Eqs. 4, 9 and 13, it is obvious that, at the boundary $S$,
\begin{equation}
\widehat{\bf t}\cdot {\bf E}_i=\widehat{\bf t}\cdot {\bf E}_t,\;
\widehat{\bf t}\cdot {\bf H}_i=\widehat{\bf t}\cdot {\bf H}_t,
\end{equation}
with $q=\exp(i{\bf k_i}\cdot {\bf u})$. Eq. 14 means that ${\bf
E}_i$ (${\bf H}_i$) and ${\bf E}_t$ (${\bf H}_t$) keep continuous
across the boundary $S$, i.e., no reflection is excited for any
propagating eigen wave incidence if Eq. 11 is satisfied. In
conclusion, Eq. 11 is the sufficient condition for a reflectionless
transformation medium in a homogenous and isotropic surrounding
medium.

Here, it should be noted the derived condition in Eq. 11 is for the
boundary, on which the transformation medium interacts with the
incident waves simultaneously. If the transformation medium has
sperate boundaries, which interact with waves in a certain sequence.
A simple example is a slab structure having two separate boundaries
for an incident wave from on side of the slav (see Ref.
\cite{Rahm2}). For such separate boundaries, the condition in Eq. 11
can be satisfied independently, i.e., different boundaries have
different $R$ and ${\bf u}$ to satisfy Eq. (11). Therefore, if part
of the boundary satisfies the condition and the waves incident
mainly upon this part, the overall scattering (reflection) could be
negligible. This well explained the mechanism of the reflectionless
design in Ref. \cite{Rahm2} in a more strict sense. In particular,
the reflectionless condition expressed in Ref. \cite{Rahm2} is that
the distances measured along the $x$, $y$ and $z$ in the
transformation medium and free space must be equal along the
boundary, where the boundary is in $y-z$ plane.This equally measured
distances in the transformation medium and free space indicates that
the coordinates can be the same before and after transformation
after a certain displacement. In fact, the distance measured along
the $x$, i.e., the direction normal to the boundary, needn't be
equal along the boundary. The reason is that the boundary is in
$y-z$ plane and a displacement in $x$ direction can always be found
to make the boundary coordinates of $x$ component be the same before
and after transformation, only if the boundary coordinates of $x$
component transform to the same value via coordinate transformation.


Furthermore, let us consider a rotation $R$ and a displacement ${\bf
u}$ operating on the coordinates $(x,y,z)$ in the original isotropic
and homogenous space before the coordinate transformation. The new
coordinates after rotation $R$ and displacement ${\bf u}$ are
denoted by $(x^{'},y^{'},z{'})$. Since the medium is isotropic and
homogenous, eigen waves are the same under the new coordinates. The
boundary coordinates of the region to be transformed is now changed
from ${\bf r_S}^{'}$ to $R{\bf r_S}^{'}+{\bf u}$ denoted by ${\bf
r_S}^{''}$, which is the same as the right term of Eq. 11. Thus, the
transformation medium can be considered as being constructed by
transformation from $(x^{'},y^{'},z{'})$ to $(q_1,q_2,q_3)$, with
the boundary coordinates keep intact before and after
transformation, i.e.,
\begin{equation}
{\bf r_s}={\bf r_s}^{''}.
\end{equation}
Therefore, in the case of an isotropic and homogenous background,
the necessary and sufficient condition expressed in Eq. 11 is
equivalent to the condition that coordinates of the boundary are
kept the same before and after transformation.

\section{Numerical examples}
To illustrate the above derived condition clearly, we present some
numerical examples here. In Fig. 2 (a), a transformation of
compressing a cylindrical region of free space with $x=2q_1$ and
$y=2q_2$ is shown. The corresponding transformation medium has the
parameters
$\widehat{\overline{\overline\epsilon}}=\widehat{\overline{\overline\mu}}=diag[1,1,4]$.
The corresponding electric field distribution for a line current
source interacting with such obtained transformation media, is
plotted in Fig. 2(b). It is clearly seen that reflections and hence
scatterings are excited, since the boundary coordinates can't be
kept the same before and after transformation by rotation and
displacement. In Fig. 3(a1) and (a2), the transformations of a
elliptic region into an elliptic annular region are shown. The outer
boundary coordinates are kept intact before and after transformation
for (a1), while the outer boundary coordinates can be kept the same
by rotation and displacement for (a2). Such transformation is
operated under free space background. The obtained transformation
media are elliptic invisibility cloaks. The electric field
distributions under a line current source for (a1) and (a2) are
plotted in (b1) and (b2), where we see no reflection is excited
since Eq. 11 can be satisfied for both cases.
\begin{figure}[htbp]
\centering
\includegraphics[width=7cm]{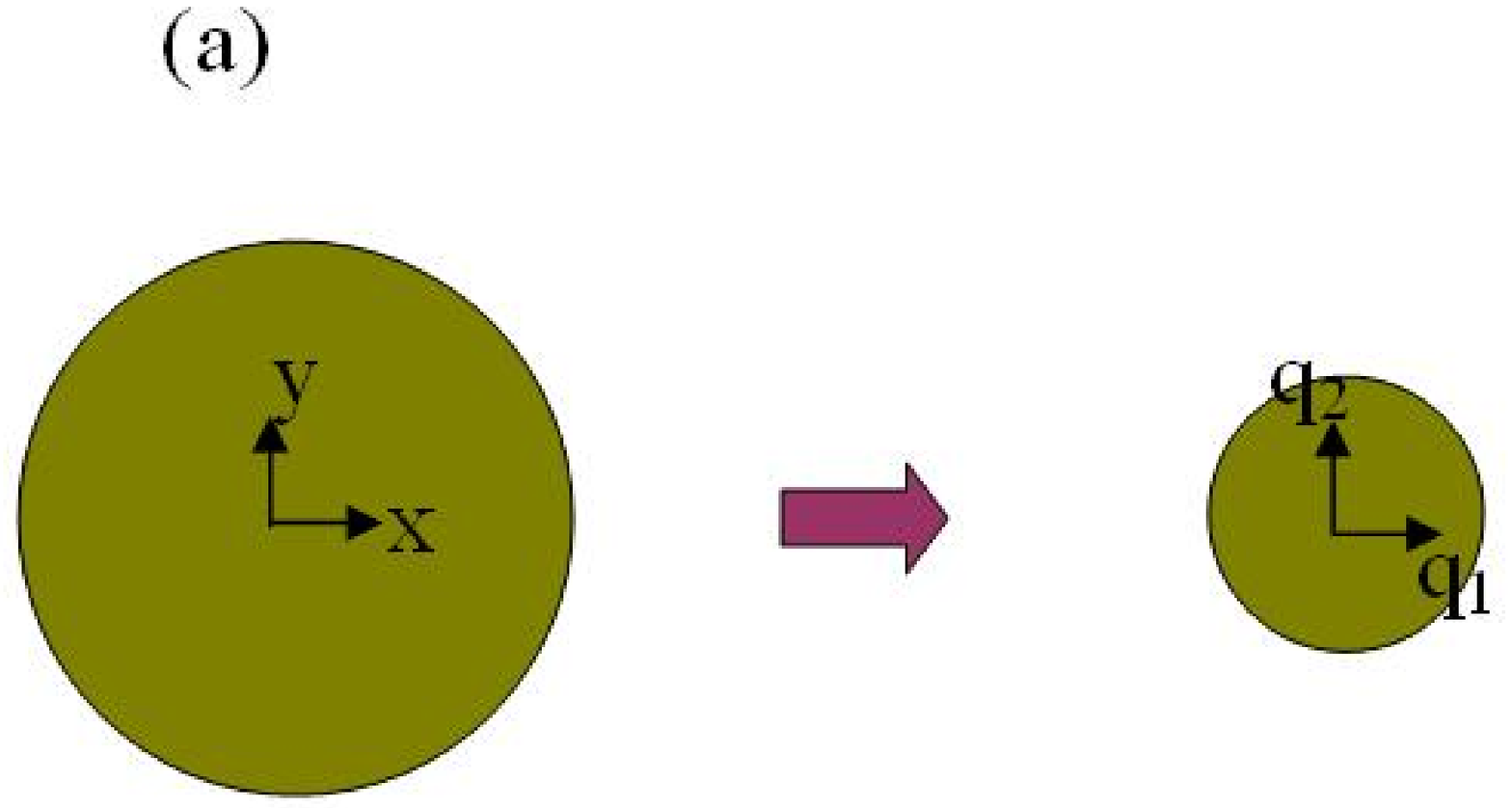}
\includegraphics[width=5cm]{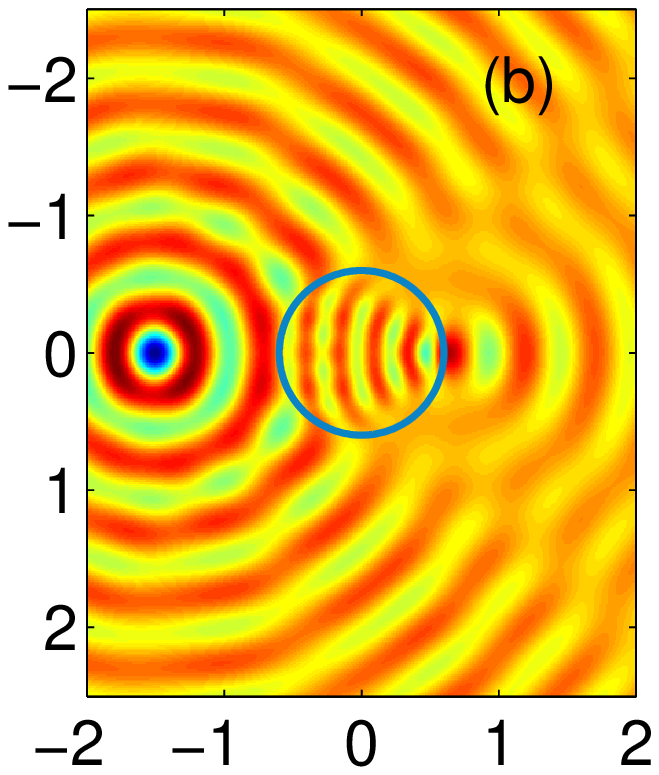}
\caption{(a) Coordinate transformation of compressing a cylindrical
region; (b) Electric field distribution for a line current source
interacting with the obtained transformation medium. The
transformation is operated under a free space background.}
\end{figure}\begin{figure}[htbp]
\centering
\includegraphics[width=6cm]{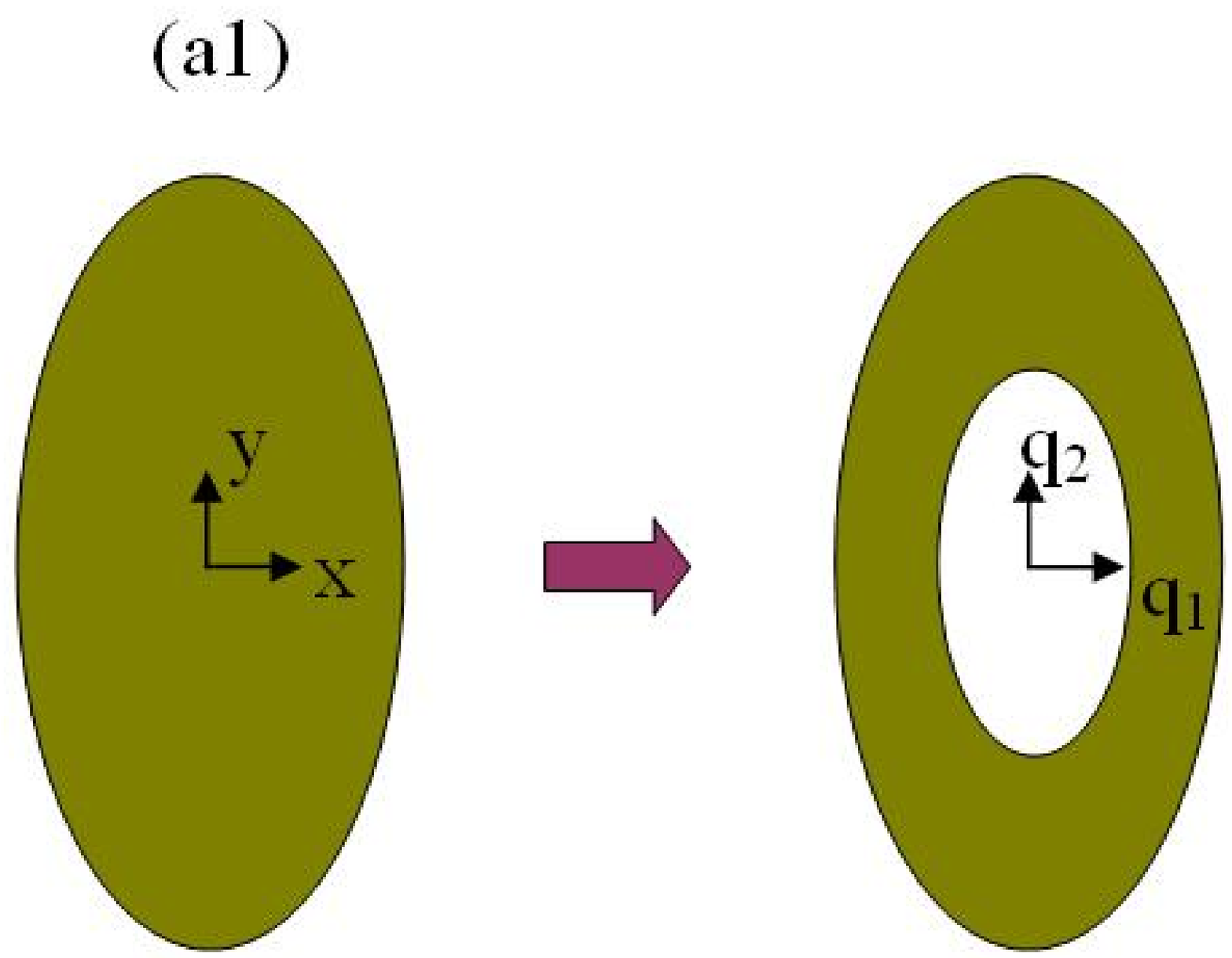}
\includegraphics[width=5cm]{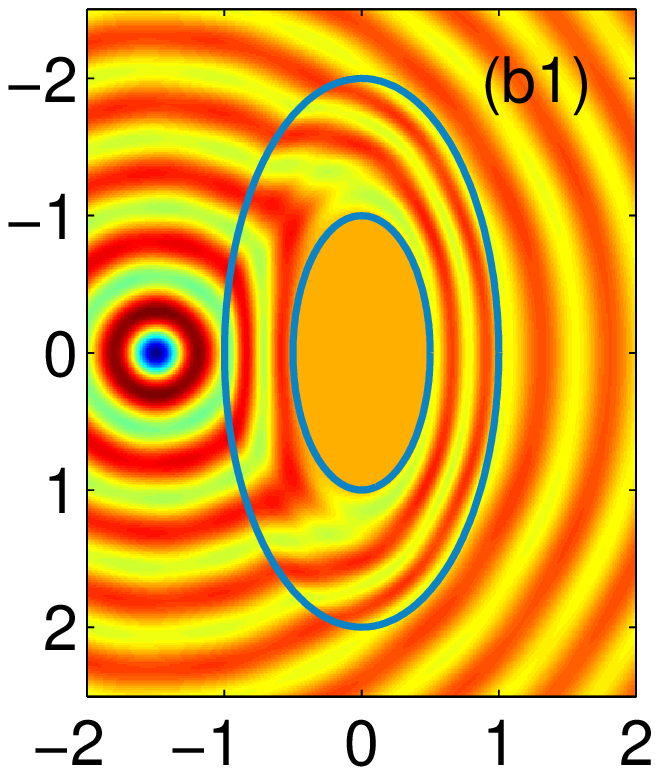}
\includegraphics[width=6cm]{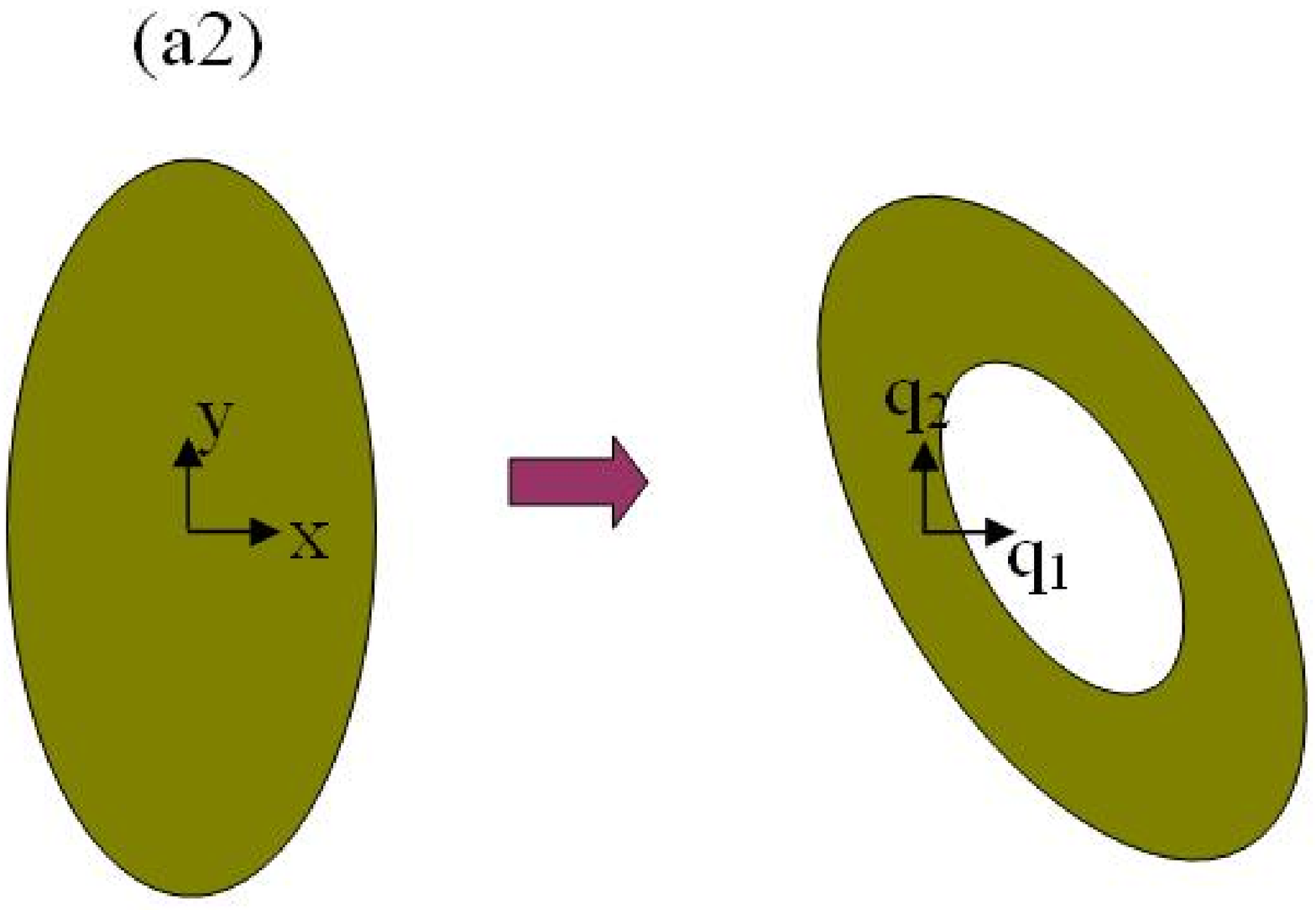}
\includegraphics[width=5cm]{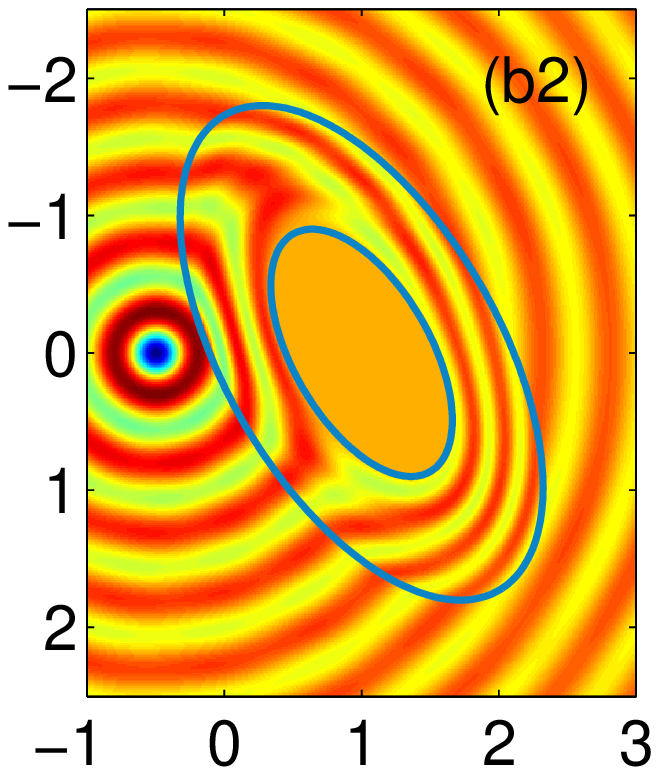}
\caption{(a1) and (a2) Coordinate transformations of a elliptic
region into a elliptic annular region; (b1) and (b2) Electric field
distribution for a line current source interacting with the
transformation media (a1) and (a2), respectively. The transformation
is operated under a free space background.}
\end{figure}\begin{figure}[htbp]
\centering
\includegraphics[width=12cm]{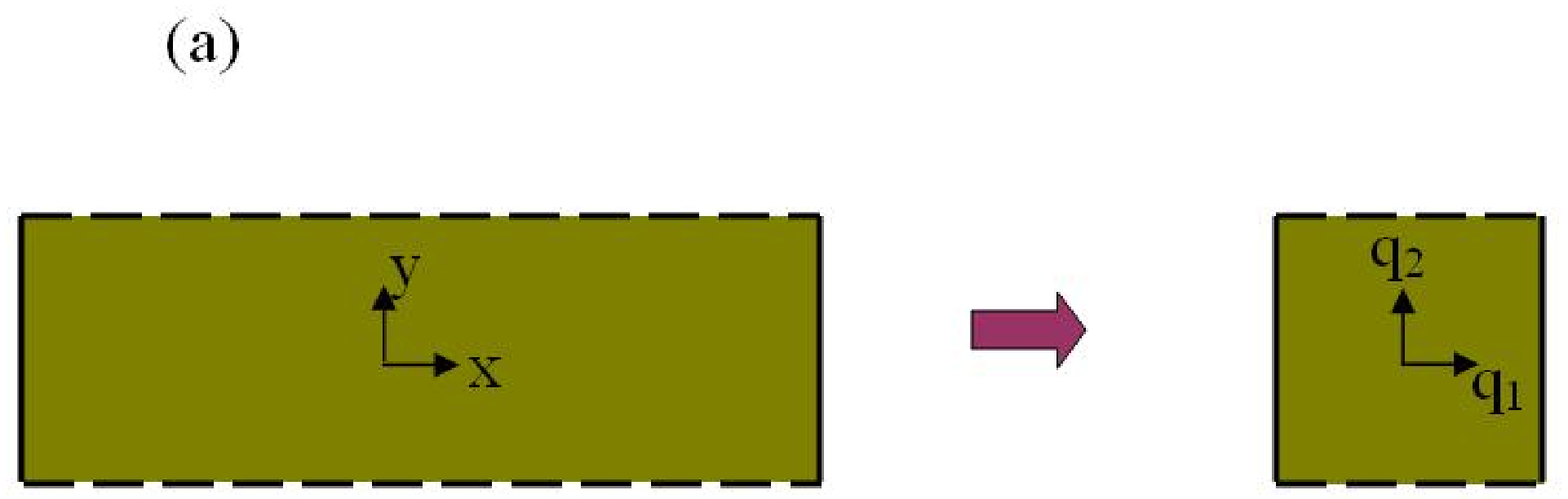}
\includegraphics[width=6cm]{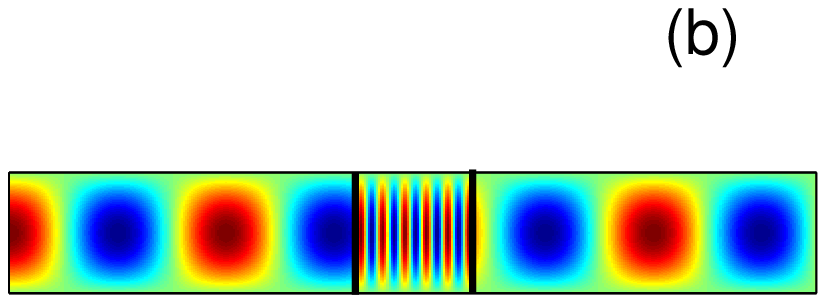}
\includegraphics[width=6cm]{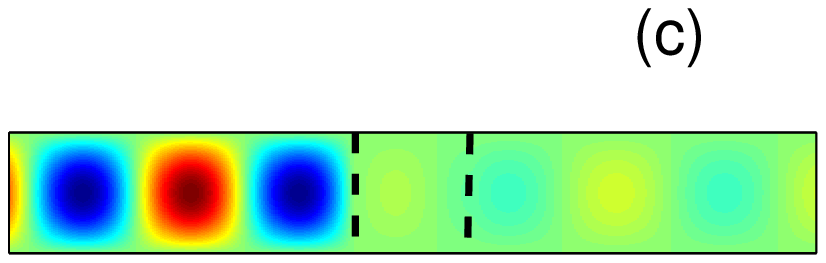}
\caption{(a) Coordinate transformation of compressing a rectangular
region into a square region. The compression is only done in $x$
direction. The left and right boundaries are denoted by solid lines,
and the up and down boundaries are denoted by dash lines. Electric
field distribution when the obtained transformation media is put in
a PEC waveguide for (b) the left and right sides of the
transformation medium interacting with the waves (c) the up and down
sides of the transformation medium interacting with the waves. The
transformation is operated under a free space background.}
\end{figure}

As discussed in the above, if one part of the boundary satisfies the
condition and the waves incident mainly upon this part, reflection
might be negligible. Consider a transformation illustrated in Fig.
4(a), where a rectangular region of free space is transformed into a
square region, by compressing the space only in $x$ direction. The
coordinates of the left and right boundaries denoted by solid lines
can be kept the same before and after transformation by
displacement. However, the coordinates of the up and down boundaries
denoted by dashed lines can't be kept the same after transformation
by rotation and displacement. Put such square transformation medium
in a PEC (Perfectly Electrical Conductor) waveguide. The wave
behaviors in the waveguide are illustrated in Fig. 4(b) and (c). In
Fig. 4(b), the waves only interact with the left and right
boundaries of the transformation medium in the propagation
direction. Thus, no reflection is observed. However, in Fig. 4(c),
the waves interact with the up and down boundaries of the
transformation medium, where reflections are excited.

\section{Discussions and conclusions}
In conclusion, we derived the necessary and sufficient condition for
a reflectionless transformation medium in an isotropic and
homogenous surrounding medium in this paper. The geometrical
expression of this condition is that the transformed coordinates of
the boundary can be the same as the original coordinates only by
rotation and displacement of the coordinates. In a general sense,
this condition is equivalent to the condition that the boundary
coordinates are kept the same before and after transformation.
Numerical simulations confirmed our findings.
\section*{Acknowledgements} This work is supported by the Swedish Foundation for Strategic Research (SSF)
through the Future Research Leaders program, the SSF Strategic
Research Center in Photon- ics, and the Swedish Research Council
(VR).


\section*{References}
\begin{thebibliography}{}
\bibitem{Pendry} Pendry J B, Schurig D, and Smith D R 2006
{\it Science} {\bf312} 1780.
\bibitem{Leonhardt} Leonhardt U 2006 {\it Science} {\bf312} 1777
\bibitem{Cummer} Cummer S A, Popa B I, Schurig D, Smith D R,
and Pendry J B 2006 {\it Phys. Rev. E.} {\bf74} 036621
\bibitem{Zolla}Zolla F, Guenneau S, Nicolet A, and Pendry J B 2007 {\it Opt. Lett.} {\bf32} 1069
\bibitem{Chen} Chen H S, Wu B I, Zhang B L, and
Kong J A 2007 {\it Phys. Rev. Lett.} {\bf99} 063903
\bibitem{Ruan} Ruan Z C, Yan M, Neff C W, and Qiu M 2007 {\it Phys. Rev. Lett.} {\bf99} 113903
\bibitem{Schurig}Schurig D, Mock J J, Justice B J, Cummer S A,
Pendry J B, Starr A F, and Smith D R 2006 {\it Science} {\bf314} 977
\bibitem{Greenleaf2} Greenleaf A, Kurylev Y, Lassas M and
Uhlmann G 2007 {\it Opt. Express} {\bf 15} 12717
\bibitem{U2} Leonhardt U 2006 {\it New J. Phys.} {\bf 8} 247
\bibitem{Rahm} Rahm M, Schurig D,
Roberts D A, Cummer S A, Smith D R and Pendry J B 2008 {\it Photon.
Nanostruct.: Fundam. Applic.} {\bf 6} 87
\bibitem{Rahm2} Rahm M, Cummer S A, Schurig D, Pendry J B, and
Smith D R 2008 {\it Phys. Rev. Lett.} {\bf 100} 063903
\bibitem{Ychen} Chen H Y and Chan C T 2007 {\it Appl. Phys. Lett.} {\bf90} 241105
\bibitem{Greenleaf} Greenleaf A, Kurylev Y, Lassas M, and
Uhlmann G 2007 {\it Phys. Rev. Lett.} {\bf 99} 183901
\bibitem{Yan} Yan W, Yan M, Ruan Z C, and Qiu M 2008
 {\it New J. Phys.} {\bf 10} 043040
 \bibitem{MY} Yan M, Yan W, and Qiu M 2008
 //www.arXiv:0804.2850[physics.optics]



\end{thebibliography}
\end{document}